\documentclass[aps,prl,twocolumn,showpacs]{revtex4}

\usepackage{graphicx}
\usepackage{amsfonts}
\usepackage{amssymb}
\usepackage{amsmath}
\usepackage{subfigure}
\usepackage{multirow}
\usepackage{tabularx}

\def\avg#1{\langle#1\rangle}

\def\beq{\begin{eqnarray}}
\def\eeq{\end{eqnarray}}

\begin{document}

\title{Biaxial nematic phases in ultracold dipolar Fermi gases}

\author{Benjamin M. Fregoso}
\author{Kai Sun}
\author{Eduardo Fradkin}
\author{Benjamin L. Lev}
\affiliation{Department of Physics, University of Illinois at Urbana-Champaign, Urbana, IL 61801-3080 USA}

\begin{abstract}
Ultracold dipolar Fermi gases represent relatively unexplored, strongly correlated systems arising from long-range and anisotropic interactions.  We demonstrate the possibility of a spontaneous symmetry breaking biaxial phase in these systems, which may be realized in, e.g., gases of ultracold polar molecules or strongly magnetic atoms.  This biaxial nematic phase is manifest in a spontaneous distortion of the Fermi surface perpendicular to the axis of polarization.  We describe these dipolar interaction induced phases using Landau Fermi liquid theory.
\end{abstract}
\date{\today}
\maketitle 

Recent experimental advances toward the creation of degenerate fermionic rovibronic ground state polar molecules, such as $^{40}$K$^{87}$Rb~\cite{Ye08}, and highly magnetic fermionic atoms, such as $^{53}$Cr~\cite{Pfau05,Gorceix06}, $^{167}$Er~\cite{McClelland06}, and $^{161}$Dy or $^{163}$Dy, present new avenues for realizing states of strongly correlated matter in ultracold atomic and molecular systems.  A degenerate dipolar Fermi gas (DDFG) is more strongly coupled and has longer range interactions than other (neutral) atomic systems, and it is therefore reasonable to expect that these systems will exhibit novel and exotic behaviors.  Recent theoretical studies have already explored BCS-like superfluidity in DDFG systems (see Ref.~\cite{Baranov2008} and references therein).

We explore the possible occurrence in homogeneously trapped DDFGs of quantum nematic phases; phenomena heretofore only observed in strongly correlated electronic systems~\cite{cooper-2002,borzi-2007,hinkov2007b} and which have been shown to occur in models based on the breakdown of Fermi liquid (FL) theory at a Pomeranchuk instability~\cite{Oganesyan2001} as well as in lattice models~\cite{kee03,Yamase2005}.  In a nematic state, the simplest example of quantum liquid crystal (QLC) order~\cite{Kivelson1998}, the fermionic gas becomes spontaneously anisotropic---without the action of an external field---while maintaining overall homogeneity.  Other inhomogeneous QLC phases, such as the smectic (stripe) phase, are both possible and likely to occur in the strong dipolar coupling regime, though are not considered here.  Quantum liquid crystal phases may also occur in highly magnetic atomic systems in optical lattices~\cite{Quintanilla2008} and in population unbalanced Fermi gases~\cite{Radzihovsky-2008}.  

We show in this work that a quantum nematic phase is possible in sufficiently strongly coupled, homogeneously trapped DDFGs. More specifically, we consider DDFGs in a fixed external polarizing field pointing along the $z$-axis.  We consider an infinite and homogeneous 3D system in which the fermionic atoms or molecules interact via the dipolar interaction that, unlike $r^{-6}$ contact interactions, has a long-range falloff $\propto r^{-3}$.  In the presence of a polarizing field, the interaction itself becomes anisotropic in real space with a $d$-wave symmetry; the dipole-dipole interaction (DDI) is repulsive or attractive depending on the spatial configuration of dipoles.  Since the DDI is explicitly anisotropic, the Hamiltonian (and ground state) of the uniformly polarized DDFG must have an explicitly broken rotational invariance, with the polarizing field providing the preferred axis.  Nevertheless, the Hamiltonian is invariant under rotations about this fixed axis, and it is this remaining symmetry that is spontaneously broken in the biaxial nematic phase.  

While DDFGs with interactions in the particle-particle channel have been previously studied~\cite{Baranov2008,You1999,Yi2001,bruun-2008}, the particle-hole (density) channel is relatively unexplored~\cite{Miyakawa2008}.  In the case of fermionic dipoles aligned by an external field, the explicit broken symmetry of the DDI caus es the coupling of even and odd orbital angular momentum channels \cite{Deb2001}.   More significantly for QLC physics, we show that an important feature of the DDI in 3D is the presence of an effective attractive particle-hole interaction in the $d$-wave angular momentum channels, which triggers a Pomeranchuk instability leading to a nematic phase of the FL.  While these phases are difficult to observe and study in electronic systems, ultracold DDFGs provide clean, experimentally realizable systems whose interaction strengths in the, e.g., $l=0$ and $l=2$, angular momentum channels can be comparable by changing external parameters such as the polarizing field and trap aspect ratio.

We now apply Landau FL theory to explore the particle-hole channel instabilities of a DDFG, and in doing so, reveal the possibility for observing a nematic state in the form of a spontaneous $xy$-symmetry breaking. The starting point of the Landau FL theory is the existence of a Fermi surface (FS) representing the ground state of the FL in the absence of quasiparticle excitations~\cite{Baym1991}. In a FL, fermionic quasiparticles have residual interactions in the forward scattering channel parametrized by a set of coupling constants, the Landau parameters of the FL. In a 3D system, such interactions can occur in any angular momentum channel. This standard picture is known to be a valid description for interacting Fermi systems such as $^3$He and most metals.  

The order parameter of a nematic state is a traceless symmetric tensor~\cite{degennes-1993}.  In a 3D interacting Fermi system, such as the DDFG, we can form a nematic order parameter as a bilinear of the Fermi fields, 
$Q_{ij}=\frac{1}{k_F^2}\psi^\dagger (\vec x) \left(\partial_i \partial_j -\frac{1}{3} \nabla^2 \delta_{ij}\right)\psi(\vec x)$,
which is both symmetric and traceless~\cite{Oganesyan2001}. Here, $i,j=x,y,z$ and $\psi(\vec x)$ is the second quantized fermion operator at position $\vec x$. The $3 \times 3$ symmetric tensor $Q_{ij}$  has two independent eigenvalues, $Q_1$ and $Q_2$. If $Q_1=Q_2$, then the nematic state is {\em uniaxial}; otherwise it is {\em biaxial}.
Since the Hamiltonian of the polarized DDFG has an explicitly broken rotational invariance (due to the polarizing field), $Q_{ij} \neq 0$ in a uniaxial state. In this case, the FS of the gas is not spherically symmetric but must instead have (at least) a uniaxial distortion: the FS is an ellipsoid oriented along the direction of polarization. In a uniformly polarized DDFG, polarization effects enter only through the anisotropy of the resulting interaction.  Hence, the uniaxial distortion of the FS must be a (generally monotonic) function of the polarization.  Indeed, Hartree-Fock (variational) calculations~\cite{Miyakawa2008}---accurate in the weak coupling regime---reveal this uniaxial FS distortion of a polarized DDFG.

Since the FS distortions are described by particle-hole condensation in specific angular momentum 
channels, the general tensor order parameter $Q_{ij}$ can be recast in terms of its angular momentum components $u_{\ell,m}$, related to the fermion occupation numbers by:
\begin{align}\label{equation1}
u_{\ell,m}=\frac{1}{V}\sum_{\mathbf{k}} Y_{\ell,m}(\hat{k})\avg{c_{\vec{k}}^\dagger c_{\vec{k}}},
\end{align}
where $c_{\vec k}$ is the Fourier transform of the Fermi field $\psi(\vec x)$, $\ell>0$, and $|m|\le \ell$; $Y_{\ell,m}(\hat k)$ are the familiar spherical harmonics.  Each order parameter $u_{\ell,m}$ measures a distortion mode of the Fermi surface.
Among them, $u_{2,0}$ and $u_{2,2}=u_{2,-2}^*$ are of main interest, as they are order parameters
for a uniaxial and biaxial nematic phase, respectively: $|u_{2,0}| \propto (Q_1+Q_2)/2$ and $|u_{2,2}| \propto |Q_1-Q_2|$.  More generally, Eq.~\eqref{equation1} represents moments of the Fermi-Dirac momentum distribution function that is anisotropic in these nematic phases.  As a consequence, in a trapped system, the density profile will distort---an effect, for the uniaxial case, seen in the bosonic dipolar condensates~\cite{Pfau07}. 

Unlike the uniaxial state, which is present in a uniformly polarized DDFG for all polarizing fields, the biaxial state is accessible only for fields of sufficient magnitude.  Whether or not this phase exists in a given dipolar gas depends sensitively on the particular Landau parameters, which is reliably calculated numerically (by intensive quantum Monte Carlo simulations) or analytically in the weak dipole limit.  The latter is defined by the condition that the dimensionless coupling constant $\lambda=nd^2/\epsilon_F$ is small, where $n$ is the particle density, $d$ is the electric or magnetic dipole moment, and $\epsilon_F=\hbar^2 (6\pi^2 n)^{2/3}/2m$ is the Fermi energy of the undistorted FS.  The Fermi temperature $T_F$ is $\epsilon_F=k_B T_F$.
Our mean field analysis predicts that a biaxial phase may arise in DDFGs for $\lambda \sim 1$. 

Consider a uniformly polarized DDFG described by the Hamiltonian:
\begin{align}
H=\sum_{\vec{k}} \epsilon({\vec{k}})c_{\vec{k}}^\dagger c_{\vec{k}}
+\frac{1}{2}\sum_{\vec{k},\vec{k}',\vec{q}}\, f(\vec{q}) \, c_{\vec{k}-\vec{q}}^\dagger c_{\vec{k}}
c_{\vec{k}'+\vec{q}}^\dagger c_{\vec{k}'},
\label{eq:Hamiltonian}
\end{align}
where
$f(\vec{q})=\frac{4\pi d^2}{3}(3\cos^2 \theta_q-1)$
is the DDI in momentum space and $\theta_q$ is the angle between $\vec{q}$ and $\hat{z}$. 
Within the mean field (Hartree-Fock) approach, the Hartree term ($\vec{q}=0$) does not contribute, 
since the average of $f(\vec{q})$ over a $4\pi$ solid angle is zero. The Fock term 
($\vec{q}=\vec{k}-\vec{k}'$) can be expanded into angular momentum channels
once the magnitudes of $\vec k$ and $\vec k'$ are set to the Fermi wavevector $k_F$ (a good approximation in the low energy theory of a Landau FL~\cite{Oganesyan2001}): 
\begin{align}
f(\vec{k}-\vec{k}')= -\sum_{\ell,m;\ell^\prime,m^\prime} f_{\ell,m;\ell',m'} Y^{*}_{\ell,m}(\hat{k})Y_{\ell',m'}(\hat{k}').
\end{align}
The minus sign originates in the Fock term.
The dimensionless coupling constants of this expansion,
$F_{\ell,m;\ell',m'}=F_{\ell',m';\ell,m}=N(0)f_{\ell,m;\ell',m'}$ are the \emph{generalized} Landau parameters, 
where $N(0)$ is the density of states at the FS. Because the 
rotational symmetry is broken by the polarization, Landau parameters $F_{\ell,m;\ell',m'}$ that mix different angular 
momentum channels enter in the Hamiltonian. For the DDI, the non-vanishing Landau parameters are
\begin{align}
F_{\ell,m;\ell,m}=&(-1)^{m}\frac{6\pi(\ell^2+\ell-3 m^2)}
{\ell(4\ell^3+8\ell^2+\ell-3)}\lambda \ \ \mbox{and}
\label{eq:Landau_parameter1}
\\
F_{\ell+2,m;\ell,m}=&(-1)^{m+1}\frac{3\pi}{2\ell^3+9\ell^2+13\ell+6}\lambda\ 
\nonumber\\
&\times\sqrt{\frac{[(\ell+1)^2-m^2][(\ell+2)^2-m^2]}{4\ell^2+12\ell+5}}.
\label{eq:Landau_parameter2}
\end{align}
More generally, the $F_{\ell,m;\ell',m'}$ will be renormalized by interactions beyond the DDI, such as the contact interaction, but
rotational symmetry and space-inversion (or time-reversal) symmetry of the system enforce the selection rules $m=m'$ and $\ell-\ell'=\textrm{even integers}$.

We now derive the free energy $F[\{ u_{\ell,m} \}]$ in order to show that there is a phase transition to a biaxial state. In the sprit of the Landau theory of phase transitions, we expand $F[\{ u_{\ell,m} \}]$ in powers of the order parameters $\{ u_{\ell,m}\}$. This procedure is accurate if the  FS distortion is small. Since the uniaxial distortion---parametrized by the order parameters $u_{2n,0}$---is present for all values of $\lambda$, this expansion is quantitatively accurate only for $\lambda \ll 1$, although it is qualitatively reasonable even for $\lambda \simeq 1$.  Additionally, the expansion in powers of the biaxial order parameters $u_{2n,m}$ ($m \neq 0$) is accurate near the critical value $\lambda_c$ for the phase transition to the biaxial state.  A perturbative (Hartree-Fock) calculation of the self energy of a DDFG at $T = 0$ yields a fermion self-energy of
$\Sigma_{HF}(\vec{k}) = \sigma(|k|)P_2(\cos\theta_k)$, where $\sigma(|k|)>0$ is a monotonic function of $|\vec k|$. This result is consistent with Eq.~\eqref{equation1} and implies that the FS is stretched along the $z$-direction with the symmetry $l=2$ for any finite $\lambda$.

To determine the structure of the free energy while keeping all order parameters obfuscates the essential physics.  We work instead with a truncated, though sufficient, set of modes: the uniaxial order parameters $u_{2,0}$ and $u_{4,0}$, and the biaxial order parameters $u_{2,\pm 2}=u_{2,\mp 2}^*$, which can be represented in terms of an amplitude $|u_{2,2}|$ and an (arbitrary) phase $\varphi_2$. 

The free-energy density of the uniaxial state, at fixed particle density and in mean-field theory up to quadratic 
order in $u_{2,0}$, has the form:
\begin{align}
\frac{F}{V}=&\frac{n}{N(0)} h_2 u_{2,0}
+\frac{1}{N(0)}m_2 u_{2,0}^2+O(u_{2,0}^4),
\label{eq:free_energy_uniaxial}
\end{align}
where $F$ is the free energy, $V$ is the volume of the system and $N(0)$ is the density of states at Fermi level. In Eq.~\eqref{eq:free_energy_uniaxial}, the term linear in $u_{2,0}$ arises from the mixing between $u_{0,0}$ (i.e., the particle density $n$) and $u_{2,0}$, and exists because $F_{2,0;0,0} \neq 0$ due to the anisotropic nature of the interaction. For $T \ll T_F$ and weak coupling $O(\lambda^3)$, the dimensionless constants $h_2$ and $m_2$ are
\begin{align}
h_2=&-\frac{2\pi^2}{7\sqrt{5}}\left(1-\frac{\pi^2 T^2}{24 T_F^2} \right) \lambda^2+O(\lambda^3), \ \mbox{and}
\\
m_2=&\frac{2\pi}{7}\lambda+\frac{3\pi^2}{35}\left(1-\frac{\pi^2 T^2}{24 T_F^2}\right) \lambda^2+O(\lambda^3),
\end{align}
with similar (but smaller) coefficients for $u_{4,0}$ and higher channels, including the mixing terms with $u_{2,0}$. 
Within this approach, a ``meta-nematic'' first order transition~\cite{Quintanilla2008,Quintanilla2008b} (i.e., a discontinuous jump in the uniaxial distortion $u_{2,0}$ driven by $\lambda$) does not occur in a 3D DDFG since $m_2>0$~\footnote{Anisotropic confinement of the gas may enhance the effective strength of the attractive interactions, leading perhaps to such a first order transition.}.

By minimizing the free energy of Eq.~\eqref {eq:free_energy_uniaxial}, the expectation value of $u_{2,0}$ (up to $O(\lambda^2)$ and $T\ll T_F$) is
\begin{align}
\frac{u_{2,0}}{n}=&\frac{\pi}{\sqrt{5}}\left(1-\frac{\pi^2 T^2}{24 T_F^2}\right) \lambda+ O(\lambda^2).
\end{align}
Hence, the FS has a uniaxial distortion in the $d$-wave channel $(2,0)$ as well as in higher even angular momentum channels, {\it e.g.\/}, $(4,0)$. 
For both Dy and KRb, the distortions in higher angular momentum channels are small, e.g., $|u_{4,0}|/|u_{2,0}| \sim 10^{-2}$.
We numerically determined the functional temperature-dependence of $u_{2,0}$ and $u_{4,0}$ to be monotonically decreasing.

We now demonstrate the existence of a phase transition to a biaxial nematic state above a critical value of the coupling constant $\lambda_c$. The leading instability to the FS occurs in the $u_{2,\pm 2}$ channels, which possess free-energy densities of\begin{align}
\frac{F}{V}=\Delta |u_{2,2}|^2+g |u_{2,2}|^4,
\label{eq:free_energy_biaxial}
\end{align}
and is independent of the phase $\varphi_2$ as expected by the $xy$ azimuthal symmetry.
In a small $\lambda$ expansion and for $T\ll T_F$, the coefficients $\Delta$ and $g$ are
\begin{align}
\Delta=&\frac{\sqrt{2}\pi^3\hbar^3}{7 m^{3/2}\sqrt{\epsilon_F}}
\lambda\Bigg[1-\frac{7\pi \lambda}{24}\left(1-\frac{\pi^2 T^2}{24 T_F^2}\right)
\Bigg], \ \mbox{and}
\label{eq:Delta}
\\
g=&\frac{761\pi^{10}\hbar^9}{254592\sqrt{2} m^{9/2}\epsilon_F^{7/2}}\left(1+\frac{1571\pi^2 T^2}{3044 T_F^2}\right)\lambda^4,
\label{eq:g}
\end{align}
where $m$ is the (effective) mass of the fermions.  We only include the leading order contributions to $\Delta$ and neglect quantitatively important effects due to, e.g., mixing between different angular momentum channels such as $(2,0)$ and $(2,2)$. Extending the present order of approximation requires either extensive numerical computations or the extrapolation of a longer series expansion~\cite{Benjamin}.

\begin{figure}[t]
\includegraphics[width=0.4\textwidth]{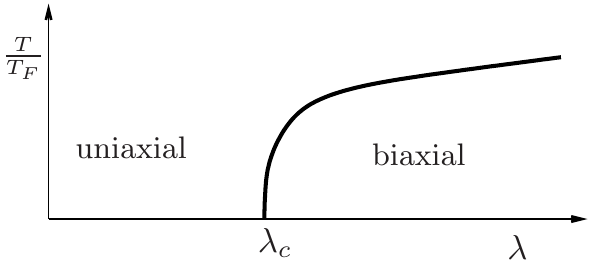}
\caption{The uniaxial-biaxial phase transition; $\lambda_c\simeq \frac{24}{7\pi}$.}
\label{fig:transition}
\vspace{-5mm}
\end{figure}

The sign of $\Delta$ dictates two different phases. For $\Delta>0$, the FL  phase with an uniaxial distortion ($u_{2,0}$)
is stable, and  $u_{2,\pm 2}=0$. For $\Delta < 0$ the system is in a biaxial nematic phase, with $|u_{2,2}|=\sqrt{|\Delta|/2g}$. For $\lambda < \lambda_c(T)$ and $\Delta>0$, the uniaxial 
nematic phase discussed above is stable. 
For $\lambda > \lambda_c(T)$  (stronger DDI), the free energy of
Eq.~\eqref{eq:free_energy_biaxial} predicts a biaxial phase. The uniaxial-biaxial phase boundary, shown in Fig.~\ref{fig:transition}, is $T/T_F= (2 \sqrt{6}/\pi) \, \sqrt{1-\lambda_c/\lambda} $, where we estimate the QCP at $\lambda_c\simeq 24/7\pi$.   The actual value of $\lambda_c$ may be larger than the mean field value predicted above due to, e.g., quantum fluctuations.  Nevertheless, signatures of this phenomenon could be within range of current experimental investigations.  

While the largest magnetic moment among the atoms (specifically, Dy and Tb with 10 $\mu_B$) provides a $\lambda_{\mbox{\footnotesize{Dy}}}$ = 0.006 at $n=10^{13}$ cm$^{-3}$, the fermionic polar molecule KRb, which has recently been laser cooled to near degeneracy, possesses a ground state electric dipole moment of 0.57 D~\cite{Ye08}.  This yields realizable $\lambda_c$'s of magnitude $[\lambda_{\mbox{\footnotesize{KRb}}},\lambda_{\mbox{\footnotesize{Dy}}}]=[0.17,0.006]$ at $n=10^{13}$ cm$^{-3}$.  A more interesting case is that of fermionic LiCs, which would possess the large dipole moment of $\sim$5.5 D~\cite{Weidemuller} and a $\lambda_{\mbox{\footnotesize{LiCs}}}\sim$ 18 for similar densities.  Other inhomogeneous phases may occur for such large values of $\lambda$. Despite rapid progress in cooling bosonic LiCs to its rovibronic ground state~\cite{WeidemullerExp}, fermionic LiCs is relatively unexplored to date.  Confinement would not substantially modify the formation of a biaxial state unless the trap is highly anisotropic.  Inhomogeneous effects in an harmonic trap induce growth of biaxial state formation from the center outward.

It is well known~\cite{Chaikin98} that 3D mean field theory fails to predict the correct critical behavior even though it yields a good qualitative picture.  Further studies using large-scale numerical simulations will be needed to refine this picture.  In particular, mean field theory does reveal that the biaxial state manifests itself as a FS distortion along two orthogonal axes, with the in-plane distortion given by $|u_{2,2}|$ and it spontaneously breaks the $SO(2)$ rotational symmetry of the system (up to in-plane rotations by $\pi$).  In the case of a DDFG confined in a very large and isotropic trap,  the correct universality class of the thermal uniaxial-biaxial transition we discuss here is the 3D classical XY model.  Like all nematics, the broken symmetry is not internal but the symmetry of spatial rotation~\cite{Nelson1981} with a fixed axis. Unlike the classical XY model---which involves an internal symmetry---the topological defects of this biaxial phase are disclination lines that cannot form closed loops because of the presence of the polarization field. Although this effect does not change the critical behavior in 3D, it does affect the behavior of the biaxial state. 

The uniaxial phase is essentially a FL state with well defined quasiparticles and an explicit static spatial anisotropy. The Fermi velocity and the speed of (zero) sound vary along the  ``meridians'' of the FS, as can be determined directly from the uniaxial distortion. We have studied the collective modes  in both phases and at the phase transition. In addition to zero sound (an $s$-wave collective mode), the uniaxial phase has three quadrupolar collective excitations, one with $L_z=0$ and two with $L_z=\pm 2$. The $L_z=0$ mode mixes with the zero sound as a result of the explicit anisotropy. 

As the phase transition to the biaxial state is approached, much as in the 2D nematic state~\cite{Oganesyan2001}, the in-plane quadrupolar collective modes become gapless and overdamped, but only when propagating in the basal $xy$-plane. This becomes a Goldstone mode in the biaxial phase but  remains gapped when propagating off-plane.  This behavior is expected since this biaxial state develops in the presence of an external symmetry breaking field: the only spontaneously broken symmetry occurs in the rotations about the direction of polarization. In contrast, a full biaxial nematic should have two gapless Goldstone modes as it has two spontaneously broken continuous symmetries~\cite{degennes-1993}.

Beyond employing destructive density measurements via light absorption to measure FS distortions, we suggest the use of polarized light scattering~\cite{degennes-1993} to detect the collective behavior of DDFGs, since in the biaxial nematic state the DDFG behaves as a birefringent medium for light propagation.  While this is a conceptually natural and direct way to access the collective mode spectrum of the nematic phase, experimentally it may be challenging to perform without destroying the nematic state.  In particular, only small scattering signals may be present due to the need for limiting the atomic excited state population while interrogating the small trap population.  An alternative method would involve measuring the anisotropy of scattering Bragg peaks;  structure factors of non-dipolar Fermi gases were recently measured using Bragg spectroscopy (see Ref.~\cite{Vale08} and citations within).

Nematic Fermi fluids---in the absence of a background lattice or at quantum criticality---are a striking example of a ``non-Fermi liquid,'' in the sense that the quasiparticles are generally broad and poorly defined~\cite{Oganesyan2001,Metzner2003}.  However, in the case of this biaxial state, the quasiparticles become broad and non-Fermi liquid-like only when they propagate in-plane due to the polarizing field.  An exciting issue is the possibility for realizing a true biaxial nematic which would exhibit exotic defects known as non-Abelian disclination lines~\cite{Poenaru-1977}.  Such defects involve twisting polarization and have skyrmionic structure. 

In the context of a DDFG, a biaxial necessarily requires that the polarization be regarded as an order parameter, i.e., ferromagnetism. Thus, the biaxial phase of a DDFG is more complex than in conventional liquid crystals. In the absence of a ferromagnetic phase of 3D DDFGs, a true biaxial phase may appear as a metastable state obtained by turning off the external polarization field.  Homogeneously trapped DDFGs are thus a natural setting to investigate the possible existence of quantum liquid crystal phases in Fermi fluids and will provide a fertile ground for explorations in non-Fermi liquid physics.

\begin{acknowledgments}
We are grateful to B. DeMarco, P. Goldbart, and N. Goldenfeld for discussions. This work was supported in part by the NSF under grants DMR 0758462 (EF) and PHY 0847469 (BLL); AFOSR FA9550-09-1-0079 (BLL); the Office of Science, U.S. Department of Energy under contracts DE-FG02-91ER45439; and through the University of Illinois Materials Research Laboratory (EF,BF,KS).
\end{acknowledgments}


\end{document}